# ACTIN: A tool to calculate stellar activity indices

**João Gomes da Silva**[1], **Pedro Figueira**[2,1], **Nuno C. Santos**[1,3], and **João P. Faria**[1,3]

**1** Instituto de Astrofísica e Ciências do Espaço, Universidade do Porto, CAUP, Rua das Estrelas, 4150-762 Porto, Portugal **2** European Southern Observatory, Alonso de Cordova 3107, Vitacura, Santiago, Chile **3** Departamento de Física e Astronomia, Faculdade de Ciências, Universidade do Porto, Rua do Campo Alegre, 4169-007 Porto, Portugal

## Summary

Magnetic activity in the atmospheres of stars produces a number of spectroscopic signatures that are visible in the shape and strength of spectral lines. These signatures can be used to access, among other things, the variability of the magnetic activity, or its influence on other parameters such as the measured radial velocity (RV). This latter is of utmost importance for the detection and characterization of planets orbiting other stars.

ACTIN is a Python program to calculate stellar activity indices. These indices can be used to study stellar activity *per se*, or to compare variations in the stellar atmospheres with RV signals. In turn, these can be used to infer if a given observed RV signal is of stellar origin or it stems from a true barycentric movement of the star. The usage of the program does not require python expertize and the indices can be customized through a configuration ASCII file.

The program reads input data either from .fits files returned by the pipelines of spectrographs, or from .rdb[1] tables. It extracts automatically the spectral data required to calculate spectral activity indices. The output is an .rdb table1, with the calculated stellar activity indices for each date (Julian Date), as well as the RV and Cross-Correlation Function profile parameters, if available. It also outputs timeseries plots of the activity indices and plots the spectral lines used to compute the indices.

The activity indices are computed based on the method described in (Gomes da Silva et al., 2011), updated with options to chose different weights and normalisations. A configuration file is provided with parameters (such as line core, bandpass, bandpass function, etc) for four widely used activity indices. These parameters can be changed and new spectral lines and indices can be added.

ACTIN was already used in research and a paper was submitted to a peer-review astrophysics journal (Delgado Mena et al., 2018, submitted).

ACTIN 1.0 comes pre-formatted for the .fits files from the HARPS and HARPS-N spectrographs, and can be easily adapted to other spectrographs such as ESPRESSO. Alternatively, since .rdb files can be read as input, ACTIN 1.0 can be easily used with any spectrum.

The code is available and will be updated on GitHub and PyPI and can be easily installed using pip.

---

[1] rdb tables are tab separated ASCII files with headers separated from the data by minus ('-') symbols with the same length as the headers.